\documentstyle[aps,prl,psfig,epsfig,multicol]{revtex}

\begin{document}

\title{Observation of a Hard Correlation Gap in Quench-Condensed Ultrathin Beryllium }

\draft

\author{E. Bielejec, J. Ruan, and Wenhao Wu}
\address{Department of Physics and Astronomy, University of Rochester,}
\address{Rochester, New York 14627}

\date{\today}

\maketitle

\begin{abstract}

We report on the tunneling density of states (DOS) in ultrathin
and strongly disordered Be films quench-condensed at 20 K. Above 5
K, the DOS shows the well-known logarithmic anomaly at the Fermi
level. Only in a narrow temperature range near 2 K is the DOS
linearly dependent on energy, as predicted by Efros and
Shklovskii. However, both the zero-bias conductance and the slope
of the linear DOS are found to decrease drastically with
decreasing temperature. Tunneling measurements at mK temperatures
have revealed conclusively that a hard correlation gap opens up in
the DOS.

\end{abstract}


\begin{multicols}{2}
It is known that electron-electron ({\em e-e}) Coulomb
interactions can drastically alter the density of states (DOS)
near the Fermi energy in disordered electronic systems.  In the
weakly disordered limit, Altshuler {\em et al.} \cite{Altshuler}
have predicted that interactions lead to a singular depletion of
the DOS with a ${\vert\epsilon\vert}^{1/2}$ dependence in three
dimensions (3D) and a ln${\vert\epsilon\vert}$ dependence
\cite{Altshuler2} in two dimensions(2D), where ${\epsilon}$ is the
energy measured from the Fermi level. These corrections have been
observed in tunneling studies of the DOS in disordered metals in
3D \cite{Dynes} and 2D \cite{White,Wu}. In the strongly insulating
regime, Efros and Shklovskii (ES) have predicted
\cite{Efros,Shklovskii} that Coulomb interactions lead to a soft
Coulomb gap in the single-particle DOS, with a vanishing DOS at
the Fermi level. This soft gap is quadratic in energy in 3D and
linear in energy in 2D. In both 2D and 3D, the Coulomb gap is
predicted \cite{Shklovskii} to lead to a variable-range hopping
resistance of R$_{\Box}$(T) = R$_{0}$ $\exp{[(T_{0}/T)^\nu]}$,
where $\nu$ = 1/2 and T is the temperature.

Although it was predicted over two decades ago, the Coulomb gap is
by no means an understood subject. The existence of the ES Coulomb
gap had mainly been inferred from transport experiments such as
glassy electronic relaxation \cite{Monroe} and hopping conduction
\cite{Shapir,Dai}. The ES Coulomb gap in 3D was directly observed
a few years ago by tunneling in Si:B \cite{Massey}. Direct
evidence for the ES Coulomb gap in 2D has been reported only
during the past year by Butko {\em et al}. \cite{Butko}, but no
temperature dependence and magnetic field dependence have been
reported. The Coulomb gap predicted by Efros and Shklovskii
\cite{Efros,Shklovskii} describes the DOS for adding an extra
electron to the ground state without allowing relaxation. Later
theoretical studies \cite{Efros2} of the Coulomb gap, taking into
consideration multi-electron processes, have found a further
reduction of the DOS near the Fermi energy, leading to a much
harder gap with effectively no states within a narrow but finite
range of energy. In fact, a change in the hopping exponent with
decreasing temperature from $\nu$ = 1/2 to $\nu$ = 1 was reported
a few years ago in Si:B \cite{Dai}, suggesting that a hard gap
might exist at low temperatures. Most recently, the Coulomb gap in
2D has become a subject of renewed interest \cite{Pastor} with the
unexpected discovery of a metal-insulator transition in the 2D
electron gas in semiconductor devices \cite{Simonian}.

In this Letter, we report tunneling measurements of the DOS in
ultrathin Be films quench-condensed near 20 K in a dilution
refrigerator. This setup makes it possible to vary film thickness
in fine steps to tune the films from the highly insulating limit
to the weakly insulating limit, which can be done {\em in situ} at
low temperatures and without exposing to air. It allows the use of
a {\em single} junction to measure, thus to compare {\em in real
units}, the densities of states from Be films of varying thickness
following successive evaporation steps. Earlier studies have
suggested \cite{Yatsuk,Semenenko,Adams} that quench-condensed Be
films are nearly amorphous. It was found \cite{Semenenko} that
superconductivity could be fully established in films as thin as
12 \AA. Although the superconducting transition temperature
T$_{c}$ of bulk Be is near 26 mK, quench-condensed Be films were
found to have surprisingly higher T$_{c}$, reaching 10 K in
thicker films \cite{Semenenko}. Scanning force microscopy studies
of our Be films, after warming up to room temperature, have found
no observable granular structure down to 1 nm.

Our Be films were thermally evaporated onto bare glass substrates
held near 20 K during evaporation. The substrates were mounted on
a rotator, extended from the mixing chamber to the bore of a 12-T
superconducting magnet. The orientation of the films with respect
to that of the magnetic field could be calibrated {\em in situ} at
low temperatures to better than 0.1$^{\circ}$ \cite{Wu2}. The
films had a multi-lead pattern for resistance measurements, with
an area of 3$\times$3 mm$^{2}$ between the neighboring leads.
Contact pads were pre-evaporated on the glass substrates. For
tunneling measurements \cite{Wolf}, one of the tunneling
electrodes was an Al film of area 1$\times$1 mm$^{2}$ and
thickness 150 \AA, evaporated at room temperature and oxidized in
air for a period of 2 hours to 2 days. After the Be films were
quench-condensed, the resistance of the resulting

\begin{figure}
\centerline{\epsfig{file=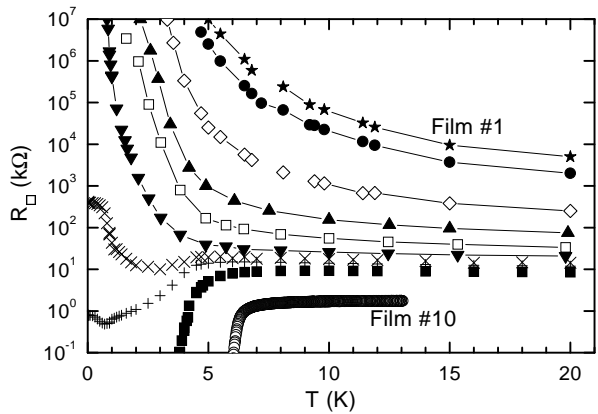,width=8.0cm}} \caption{Curves
of film sheet resistance as a function of temperature measured on
one film section following a series of deposition steps to
increase film thickness. For curves from top to bottom, we label
them as Film $\#$1 to Film $\#$10, respectively. The thickness for
these films changed from 4.6 {\AA} to 15.5 \AA. }\label{Figure 1}
\end{figure}

\noindent Be/Al$_{2}$O$_{3}$/Al junctions were found to vary from
5 k$\Omega$ to 1 M$\Omega$ at 20 K. All the tunneling data and
nearly all the film resistance data shown below were obtained by
calculating the numerical derivative of the 4-terminal dc I-V
curves measured using two Keithley 617 electrometers.

Figure 1 shows the temperature dependence of the film sheet
resistance, R$_{\Box}$, measured on one film section deposited on
a bare glass substrate following successive deposition steps to
increase film thickness. The film changed its behavior from
insulating to superconducting when R$_{\Box}$ at 20 K was reduced
to below 10 k$\Omega$/$\Box$ as the film thickness was increased.
Film $\#$10 in Fig. 1, which was superconducting with T$_{c}$
$\sim$ 6 K, had a critical field H$_{c}$ well above the 10-T field
our magnet could reach at 4.2 K. Using the spin-paramagnetic limit
\cite{Fulde}, we estimate that for this film the upper-bound of
H$_{c}$ is $\sqrt{2}\Delta$/g$\mu_{B}$ $\approx$ 13.5 T, where g
$\approx$ 2 is the Land$\acute{e}$ g-factor, $\mu_{B}$ is the Bohr
magneton, and $\Delta$ $\approx$ 0.92 mV is the superconducting
gap (see below). Early studies \cite{Lazarev} estimated that the
critical field was 18 $\sim$ 20 T in quench-condensed Be films of
T$_{c}$ = 8 $\sim$ 10 K, suggesting that these films were highly
disordered with a very short penetration depth.

For tunneling DOS, we measure the dc tunneling I-V from which the
tunneling conductance, G = dI/dV, is calculated. When thermal
broadening is unimportant, G is simply the product of the
tunneling probability and the densities of states of the two
electrodes. The quality of our junctions could be tested when the
Be films became thick enough so that they were superconducting,
such as Film $\#$10 in Fig. 1. Our tunneling studies at 100 mK on
Film $\#$10 found that the combined superconducting gaps of Be and
Al was 1.20 mV. In a 1.5-T perpendicular magnetic field,
H$_{\perp}$, which was strong enough to suppress superconductivity
in the Al electrode and yet was too weak to produce any measurable
effect on the gap value of Film $\#$10, we measured an energy gap
of 0.92 mV for Film $\#$10. Given that the T$_{c}$ of Film $\#$10
was near 6 K, this gap of 0.92 mV led to 2$\Delta$ $\approx$
3.7k$_{B}$T$_{c}$. We note that, for the zero-field data presented
below, we do not attempt to extract the DOS profile of the
superconducting Al electrode from the measured tunneling
conductance. We treat the Al electrode as having a flat DOS
because, as will be shown later, the energy scale associated with
the DOS structures of the highly insulating Be films is far larger
than the gap energy ($\sim$ 0.28 mV) of the superconducting Al
electrode. Finally, the resistance of our junctions was always at
least three orders of magnitude larger than the resistance of the
Be film sections forming the leads for the junctions. As a result,
the tunneling current produced a negligible voltage drop on the
leads, ensuring that the measured voltage was the bias voltage
across the junctions. We point out that the sheet resistance of
the portions of the Be film deposited on top of the
Al$_{2}$O$_{3}$/Al layers, which formed the junctions, could be
different from the rest of the film deposited on the bare glass
substrates. Thus one should be cautious when comparing the
tunneling data presented below with the film sheet resistance
shown in Fig. 1. The goal of this Letter is to report the
tunneling DOS in quench-condensed ultrathin Be films, showing the
evolution from a logarithmic anomaly to the ES linear Coulomb gap,
and the eventual opening up of a hard gap. Simultaneous tunneling
and resistance measurements in films deposited on exactly the same
under-coating or substrates will be carried out in the future.

In Fig. 2, we plot the tunneling conductance measured at three
temperatures of one of the junctions on Film $\#$6 in Fig. 1. The
data near 3 K and 2 K in Fig. 2 appear to show the ES linear
Coulomb gap \cite{Efros,Shklovskii} in 2D. However, with
decreasing temperature, the slope of the linear DOS decreased
sharply and the tunneling conductance at zero-bias dropped toward
zero. Thus the DOS can not be described simply by the
temperature-independent ES linear Coulomb gap. We point out that
the resistance data measured from insulating films (films $\#$1 to
$\#$6) in Fig. 1 can not be fitted to a simple hopping law with
$\nu$ = 1/4, 1/3, 1/2, or 1. This is probably not surprising,
given that the measured tunneling DOS depends strongly on
temperature. In contrast, the ES hopping law with $\nu$ = 1/2 is
derived for a temperature-independent soft Coulomb gap. We note
that a significant temperature dependence of the tunneling DOS has
been reported before in both 2D \cite{Wu} and 3D \cite{Massey}.
The temperature dependence seen in Fig. 2 is likely due to the
following reason. At low temperatures, electrons are strongly
localized with a very low diffusivity, resulting in a very poor
screening capability. This is the regime where the ES Coulomb gap
is expected. As the temperature is raised, the diffusivity
increases and the screening capability is improved. In fact, the
tunneling conductance at higher temperatures, such as 5.74 K, is
logarithmically in energy, as predicted in the weakly disordered
limit \cite{Altshuler}.

\begin{figure} \centerline{\epsfig{file=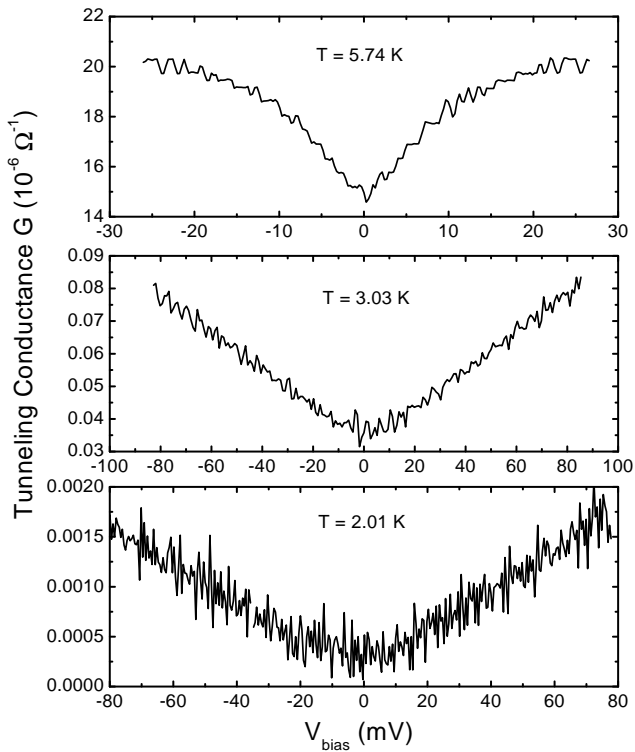,width=8.5cm}}
\caption{Tunneling conductance, G, obtained from one junction on
Film $\#$6 in Fig. 1, as a function of V$_{bias}$ at three
temperatures, showing the depletion of the DOS by orders of
magnitude as temperature was lowered. The noise seen in the data
resulted from calculating the numerical derivative of the dc I-V
curves.}
\label{Figure 2}
\end{figure}

Film $\#$6 from Fig 1 was so insulating that tunneling
measurements could not be made on this film much below 2 K,
because at low temperatures such films formed highly resistive
leads for the junctions. Earlier reports [11,12] have also
discussed technical difficulties in tunneling experiments
associated with high lead resistance. Tunneling measurements at
much lower temperatures were possible only on slightly thicker
films which were much more conducting, such as films $\#$7 and
$\#$8 in Fig. 1. The low resistance of such films ensured that the
measured voltage was across the junctions, as we discussed
earlier. In the inset to Fig. 3, we plot one dc tunneling I-V
measured at 30 mK on Film $\#$8. We see clearly a wide hard gap of
about 30 mV in the DOS. In the main part of Fig. 3, we plot on
both linear and log-log scales the tunneling conductance curves
measured at a number of temperatures for T $\lesssim$ 3.3 K. Above
5 K, the tunneling conductance also had a logarithmic anomaly.
Apart from a sharp dip near zero-bias, the tunneling conductance
near 2 K showed a V-shaped linear energy dependence. The DOS
depleted further by orders of magnitude with decreasing
temperature and developed a temperature-independent hard gap below
500 mK.

Films $\#$7 and $\#$8 in Fig. 1 were quite conducting due to the
existence of strong superconducting fluctuations.

\begin{figure}
\centerline{\epsfig{file=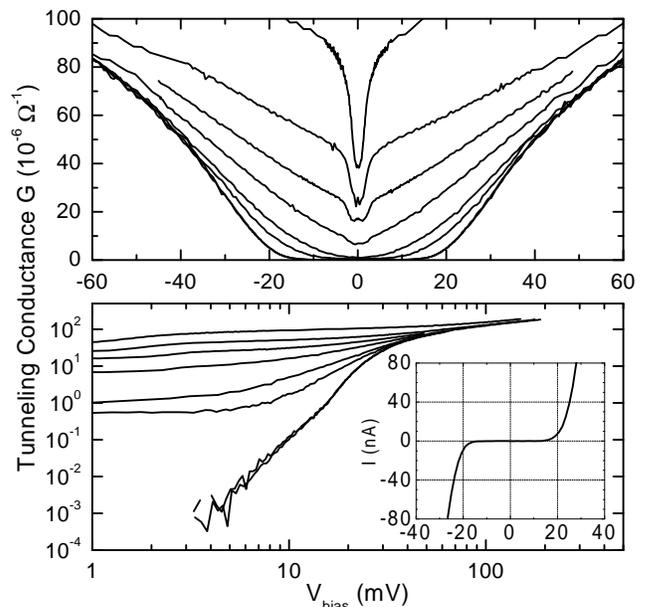,width=8.5cm}} \caption{Main
figures show the tunneling conductance, G, as a function of bias
voltage at various temperatures measured on Film $\#$8 in Fig. 1,
on both linear (top plot) and log-log (bottom plot) scales. In
both plots, curves from top to bottom were measured at 3.30 K,
2.80 K, 2.50 K, 2.10 K, 1.40 K, 1.00 K, 500 mK, and 30 mK,
respectively. The 500 mK and 30 mK curves fall on top of each
other and are nearly indistinguishable. Inset: A dc tunneling I-V
measured at 30 mK on Film $\#$8, showing a hard gap of about 30
mV.}
\label{Figure 3}
\end{figure}

\noindent Nevertheless, we now argue that this hard gap is not a
result of these superconducting fluctuations; rather, we believe
that the hard gap is a manifestation of localization-enhanced
interaction effects in strongly disordered films, and would be
found in all our insulating samples if only we were able to
measure tunneling in extremely insulating thinner films such as
Film $\#$6. First, it seems unlikely that the hard gap originates
from superconductivity since the width of the hard gap, 30 $\sim$
40 mV, is so much larger than the superconducting gap, $\Delta$
$\approx$ 0.92 mV, measured on Film $\#$10. Next, we found that
the width of the hard gap was reduced from 40 mV in Film $\#$7 to
30 mV in Film $\#$8. Although our crystal thickness monitor was
not able to resolve the difference in thickness between these two
films, Film $\#$8 was produce by an additional evaporation step
upon Film $\#$7, and so is slightly thicker. Hence, we have found
a correlation between a thinner film and a wider hard gap.
Finally, for a fixed film thickness, we found that the hard gap
broadened significantly with increasing perpendicular magnetic
fields, H$_{\perp}$, in the low-field regime, as shown in Fig. 4.
The hard gap was insensitive to a parallel magnetic field in the
same field range. Such highly anisotropic behavior is most likely
an orbital effect rather than a spin effect. Since a perpendicular
magnetic field effectively localizes the electrons in
2D\cite{Eisenstein}, we believe that the field-broadening of the
hard

\begin{figure}
\centerline{\epsfig{file=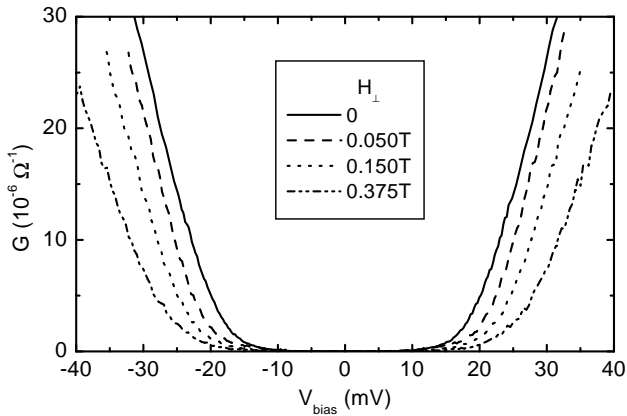,width=8.5cm}}
\caption{Tunneling conductance, G, versus the bias voltage
measured at 30 mK on Film $\#$8 in Fig. 1, showing the broadening
of the hard gap with increasing H$_{\perp}$. }
\label{Figure 4}
\end{figure}

\noindent gap in Fig. 4 indicates a correlation between the hard
gap and localization. In fact, correlation gaps induced by the
localization effect of a strong perpendicular magnetic field have
been observed before\cite{Eisenstein} in tunneling studies of the
2D electron gas in semiconductor quantum wells even in the limit
of insignificant disorder.

In conclusion, our tunneling studies of quench-condensed ultrathin
Be films have revealed a strong dependence of the DOS on film
thickness, temperature, and magnetic field. We have seen that the
DOS evolves from a logarithmic anomaly to the ES linear Coulomb
gap near 5 K. At mK temperatures, a hard correlation gap as wide
as 30 $\sim$ 40 mV emerges at the Fermi level. As the films become
thicker and less insulating, the hard gap narrows. Such behavior
has never been observed before. We argue that this hard
correlation gap results from the combination of localization and
{\em e-e} Coulomb interactions, and that the hard gap should also
exist in the strongly insulating Be films on which we were not
able to perform tunneling experiments.

We gratefully acknowledge numerous invaluable discussions with S.
Teitel, Y. Shapir, Y. Gao, and P. Adams. We thank S. Zorba and Y.
Gao who performed scanning force microscopy studies of our
quench-condensed Be films.
\end{multicols}
\end{document}